\documentclass[%
 reprint,
 amsmath,amssymb,
 aps,
]{revtex4-1}

\usepackage{latexsym}
\usepackage{textcomp}
\usepackage{graphics}
\usepackage{graphicx}

\usepackage{amsmath}

\begin{document}

%%\twocolumn[ %% activate for two-column option

\title{Bottle microresonator with actively stabilized evanescent coupling}

%% For REVTeX it is possible to automate superscript and e-mail callouts with the superscriptaddress option; see REVTeX4 documentation.

\author{C. Junge,$^{1,2}$ S. Nickel,$^2$ D. O'Shea,$^{1,2}$ and A. Rauschenbeutel$^{1,2,*}$}

\affiliation{
$^1$Vienna Center for Quantum Science and Technology, \\ TU Wien - Atominstitut, Stadionallee 2, 1020 Wien, Austria
\\
$^2$Institut f\"{u}r Physik, Johannes Gutenberg-Universit\"{a}t Mainz, 55099 Mainz, Germany \\
$^*$Corresponding author: arno.rauschenbeutel@ati.ac.at
}

\date{\today}

\begin{abstract}The evanescent coupling of light between a whispering-gallery-mode bottle microresonator and a sub-wavelength-diameter coupling fiber is actively stabilized by means of a Pound-Drever--Hall technique. We demonstrate the stabilization of a critically coupled resonator with a control bandwidth of 0.1~Hz, yielding a residual transmission of $(9 \pm 3) \times 10^{-3}$ for more than an hour. Simultaneously, the frequency of the resonator mode is actively stabilized.
\end{abstract}
%\ocis{230.5750, 230.3990, 140.3945, 060.1810.}

 %%] %% activate for two-column option
\maketitle

\begin{figure}[b]
\centerline{\includegraphics[width=7.5cm]{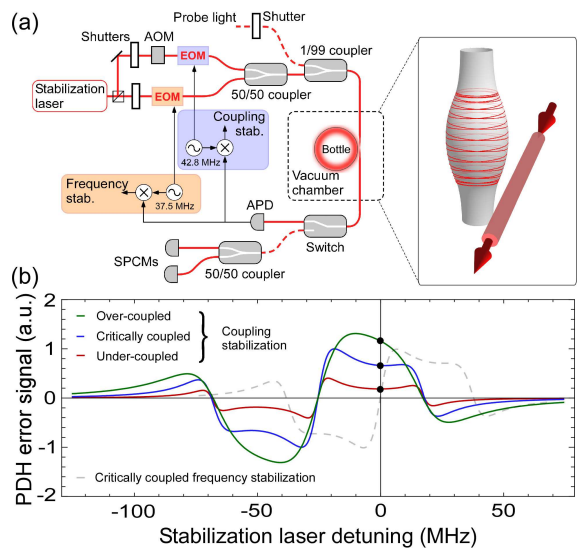}}
\caption{(a) Experimental setup. The inset schematically shows the bottle resonator and the coupling fiber; (b) PDH error signal, used for coupling stabilization, as a function of the laser--resonator detuning. The curves have been calculated for three different coupling strengths with $\Delta \nu_{\rm int}=3.2$~MHz and $\nu_{\rm mod,c}=42.8$~MHz. Dashed line: Calculated PDH signal used for frequency stabilization  with $\nu_{\rm mod,f}=37.5$~MHz.}\label{fig1}
\end{figure}

\noindent
Microresonators are key elements in a wide range of applications such as microlasers, cavity quantum electro dynamics (CQED), cavity opto-mechanics, and optical signal processing \cite{Vahala:2003aa,Ilchenko:2006,Kippenberg:2008,Notomi:2010}. In this context, whispering-gallery-mode (WGM) microresonators have attained special interest since they combine ultra-high quality factors and small mode volumes \cite{Kippenberg:2004aa} with the ability to evanescently couple light into and out of the resonator with very high efficiency using evanescent couplers such as prisms, angle-polished fibers and micro-tapered optical fibers \cite{Knight:1997aa,Spillane2003a,Matsko2006}. WGM resonators are monolithic dielectric structures that confine light near the resonator surface by continuous total internal reflection \cite{Matsko2006}. They have been realized in several different designs, such as microspheres, microtoroids, microdisks \cite{Vahala:2003aa,Matsko2006}, and, recently, bottle microresonators \cite{Sumetsky04,Louyer2005,SenthilMurugan:09,Pollinger2009} (see inset to Fig.~\ref{fig1}~(a)). While being conceptually similar to traditional WGM resonators and sharing many properties, bottle microresonators have the additional advantage of combining the tunability typical to Fabry-P{\'e}rot resonators. Light is coupled into such resonators by matching the phase and frequency of the field of the evanescent coupler with the WGM while the coupling strength is set by their spatial overlap \cite{Knight:1997aa}. This coupling strength determines the intracavity intensity for a given input power. The latter is maximized in the case of so-called critical coupling, i.e., when the losses induced by the evanescent coupler equal the intrinsic round trip losses in the cavity. In order to maintain a stable intracavity intensity, the gap between an evanescent coupler and a WGM microresonator must remain stable within a few tens of nanometers. So far, evanescent coupling has mostly relied on the passive stability of the setup. In practice, however, long-term drifts on the nanometer scale will occur.

At critical coupling, the intensity of the light field transmitted through the evanescent coupler past the resonator is not a monotonous function of the coupling strength but goes through a minimum \cite{Spillane2003a}. Thus, the transmitted intensity cannot be directly used as an error signal for stabilizing the coupling gap. Instead, a modulation technique can be employed in order to gain information about the sign of the change in coupling strength.

In this letter, we present a novel method, based on the Pound-Drever--Hall (PDH) modulation technique \cite{Drever:1983aa}, to actively stabilize the evanescent coupling between a bottle microresonator and the sub-wavelength-diameter waist of a tapered optical fiber (coupling fiber). The idea of the scheme is to use an off-resonant PDH signal to sense the linewidth broadening of the bottle mode which is proportional to the coupling strength. This scheme avoids any mechanical modulation of the coupling gap size \cite{Wilcut:09} and its bandwidth can in principle reach the dynamical limits of the mechanical coupling setup.

The amplitude of the PDH error signal for light of frequency $\nu$ is given by the relation \cite{Black2001}
\begin{equation}\label{eqn:PDH}
   \epsilon(\nu) \propto \text{Im}\{t(\nu) t^{*}(\nu+\nu_{\rm mod})-t^{*}(\nu) t^{}(\nu-\nu_{\rm mod})\},
\end{equation}
where $\nu_{\rm mod}$ is the phase modulation frequency which also determines the spectral separation of the carrier and the sidebands. For a WGM resonator, the amplitude transmission through the coupling fiber, $t(\nu)$, also depends on the coupling gap $x$ and is given by \cite{Haus:84aa}
\begin{equation}\label{eqn:T}
    t(x,\nu) = \frac{\Delta \nu(x)-\Delta \nu_{\rm int}- 2 i (\nu-\nu_{0})}{\Delta \nu(x)+\Delta \nu_{\rm int}+ 2 i (\nu-\nu_{0})}.
\end{equation}
Here, $\Delta \nu_{\rm int}$ and $\nu_{0}$ are the intrinsic linewidth (FWHM) and the resonance frequency of the resonator mode, respectively. The coupling fiber induces a gap-dependent linewidth broadening $\Delta \nu(x)$ which scales as $\exp(-x/ \gamma_{0})$, where $\gamma_{0}$ is the evanescent field decay length.

Figure~\ref{fig1}~(b) shows the calculated PDH signal for different coupling regimes. For decreasing coupling gaps, the linewidth increases and there is a corresponding increase of the off-resonant PDH signal (solid black circles) which therefore behaves monotonically for a large range of gap sizes. In order to generate the off-resonant PDH signal, a constant resonator--laser detuning is required. The experimental setup, shown in Fig.~\ref{fig1}~(a), therefore uses two laser beams derived from an external cavity diode laser (50~kHz linewidth) which are phase modulated using electro-optic modulators (EOMs). For one laser beam ($\nu_{\rm mod,f}=37.5$~MHz), the carrier is resonant with the bottle mode and is used for stabilizing the frequency of the resonator \cite{OShea2010}. Simultaneously, a second, laser beam with a constant carrier frequency detuning of 20--25~MHz relative to the resonance frequency of the bottle mode is used to create the off-resonant PDH signal for the coupling stabilization ($\nu_{\rm mod,c}=42.8$~MHz). Both beams are sent through the coupling fiber and the light transmitted past the resonator is detected with an avalanche photodiode (APD). The APD signal is split and mixed with the modulation frequencies $\nu_{\rm mod,f}$ and $\nu_{\rm mod,c}$, respectively, to create the two PDH signals.

The bottle microresonator has a diameter of 38~$\mu$m and is made from a silica glass fiber in a heat-and-pull-process as described in \cite{Pollinger2009}. In this work, we excite bottle modes with linewidths of \mbox{8--15~MHz} at critical coupling, corresponding to intrinsic quality factors of \mbox{$Q_{\rm int}=0.5$--$1\times10^8$}. We tune the bottle mode frequency via mechanical strain, applied to the glass fiber carrying the resonator by means of a shear piezo stack. The coupling fiber is mounted on a positioning stage that enables sub-nanometer control of the coupling gap. The coupling setup is placed in a vacuum chamber with a background pressure of $1\times10^{-9}$~mbar, compatible with the requirements of a cold-atom CQED experiment \cite{oshea:2011}. We note, however, that our scheme should work equally well under ambient conditions.

\begin{figure}[t]
\centerline{\includegraphics[width=7.5cm]{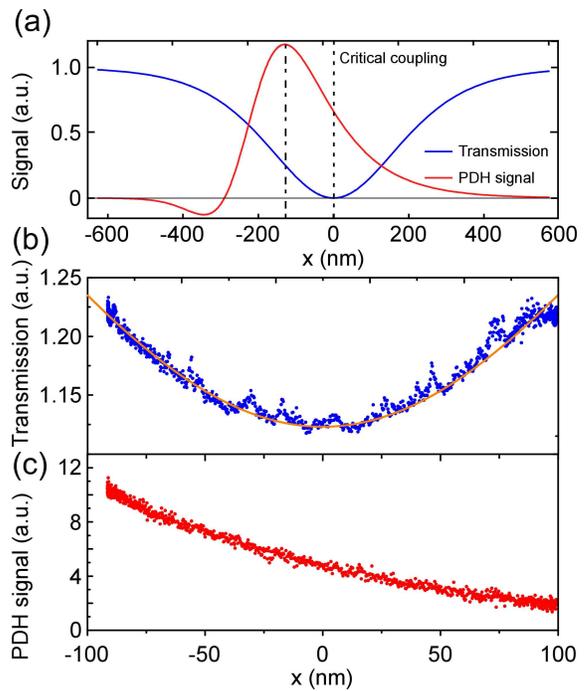}}
\caption{(a) Plot of the intensity transmission $|t(x,\nu)|^2$ and of the off-resonant PDH signal as a function of the coupling gap according to Eqns.~(\ref{eqn:PDH}) and (\ref{eqn:T}). The zero on the $x$-axis corresponds to the coupling gap at critical coupling. The experimental data in (b) and (c) show the transmission and the off-resonant PDH signal around the critical coupling point. The solid line in (b) is a fit of $|t(x,\nu)|^2$ according to Eqn.~(\ref{eqn:T}) including an offset.}\label{fig2}
\end{figure}

Figure~\ref{fig2}~(a) shows a plot of the calculated intensity transmission $|t(x,\nu)|^2$ and of the calculated off-resonant PDH signal as a function of the coupling gap for our experimental parameters. As discussed above, $|t(x,\nu)|^2$, vanishes at critical coupling and all light is dissipated in the resonator. The off-resonant PDH signal has the desired monotone behavior for gap variations around critical coupling and is therefore a suitable error signal for coupling stabilization. For decreasing coupling gaps, the mode linewidth can exceed the modulation frequency. If the detuning from resonance is half the modulation frequency, $\nu-\nu_{0}=\nu_{\rm mod,c}/2$, this results in a rolloff of the PDH signal for a coupling gap $x_{\rm rolloff}$ (see dash-dotted line in Fig.~\ref{fig2}~(a)), determined by the condition $\Delta \nu(x_{\rm rolloff})+\Delta \nu_{\rm int} = \nu_{\rm mod,c}$. Thus, for a given $Q$-factor, $x_{\rm rolloff}$ can be shifted over a large range of values by appropriate choice of $\nu_{\rm mod,c}$.

\begin{figure}[t]
\centerline{\includegraphics[width=7.5cm]{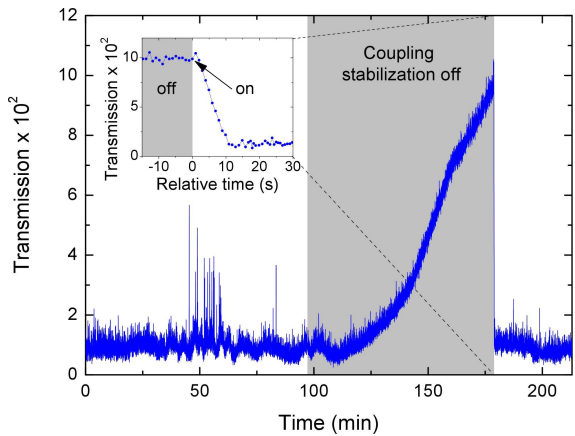}}
\caption{Long-term measurement of the intensity transmission characterizing the performance of the coupling stabilization at critical coupling. The stabilization is switched off and on again at 90~min and 160~min, respectively. Inset: Zoom of the recapturing process.}\label{fig3}
\end{figure}

In order to experimentally verify our simulations, we scanned the coupling gap around the critical coupling value in open-loop mode and recorded simultaneously the intensity transmission detected by the APD and the off-resonant PDH signal, see Fig.~\ref{fig2}~(b),~(c). The transmission and the off-resonant PDH signal are in good qualitative agreement with the theoretical predictions in Fig.~\ref{fig2}~(a). In particular, the PDH signal is monotonous around critical coupling.

We verify the correct operation of the stabilization scheme by using a lock--probe method: The system is cyclically stabilized for one second and subsequently, the transmission is probed for 5~ms with a weak (pW), resonant laser beam (see Fig.~\ref{fig1}~(a)). During the probing, the frequency and coupling stabilization beams (with about 20~nW in each beam) are turned off with shutters and the probe transmission is detected with two single photon counting modules (SPCMs).

Figure~\ref{fig3} shows the corresponding probe transmission, recorded over several hours. It can be seen that the system remains stabilized at the critical coupling point with a very low average intensity transmission of $9 \times 10^{-3}$ while the standard deviation is $3 \times 10^{-3}$. After 90~minutes, the coupling stabilization is turned off while the frequency stabilization is still operating. In the following, drifts of the coupling gap cause the transmission to change, reaching a value of 0.1 after 70~min. At that point, the coupling stabilization is switched on again and the stabilization scheme successfully restores the critical coupling condition. The inset of Fig.~\ref{fig3} shows a zoom of the recapturing process. From this signal, we estimate the bandwidth for the stabilization technique to be 0.1--0.2~Hz. This value is limited by the time constant of a low pass filter that has been inserted into the control loop in order to avoid the excitation of mechanical resonances of the coupling setup. The temporary transmission fluctuations around 50~min occur on a timescale of 1--7~s and are assigned to mechanical vibrations of the coupling setup probably from seismic, thermal, or acoustic noise at frequencies exceeding the stabilization bandwidth.

From the transmission noise in Fig.~\ref{fig3}, we furthermore determine the fluctuations of the coupling gap. At critical coupling, these fluctuations cause asymmetric noise that can be differentiated from noise sources that yield a symmetric distribution like, e.g., detector noise. From this analysis, we estimate the rms-fluctuation of the gap to be $\pm9$~nm in the frequency range up to 200~Hz.

In conclusion, our method holds great potential for applications involving evanescently coupled WGM microresonators if a high coupling stability is required over an extended period of time, especially in the case of critical coupling. Such applications range, e.g., from optical signal processing \cite{Ilchenko:2006,Pollinger:2010} and CQED with WGM resonators \cite{Aoki:2006aa} to single molecule detection with cavities in bio-sensing \cite{Armani2007}. Moreover, our technique can be equally well applied to other evanescent coupling schemes \cite{Matsko2006} and other types of evanescently coupled cavities, such as, e.g., photonic crystals \cite{Srinivasan2004}.

We gratefully acknowledged financial support by the DFG (Research Unit 557), the Volkswagen Foundation (Lichtenberg Professorship), and the ESF (EURYI).

%\pagebreak

%\bibliographystyle{osajnl}
%\bibliography{references}

\end{document}